\documentclass[fleqn,usenatbib]{mnras}

\usepackage{newtxtext,newtxmath}

\usepackage[T1]{fontenc}
\DeclareRobustCommand{\VAN}[3]{#2}
\let\VANthebibliography\thebibliography
\def\thebibliography{\DeclareRobustCommand{\VAN}[3]{##3}\VANthebibliography}

\usepackage{graphicx}	
\usepackage{amsmath}	

\title[Interior models of KIC 7747078]{Is there a unique asteroseismic interior model for the solar-like oscillating KIC 7747078?}

\author[S. Örtel, M. Yıldız and Z. Çelik Orhan]{
Sibel Örtel$^{1,2}$\thanks{E-mail: sibel.ortel@gmail.com} \thanks{ORCID: 0000-0001-5759-7790},
Mutlu Yıldız$^{1}$\thanks{ORCID: 0000-0002-7772-7641},
Zeynep Çelik Orhan$^{1}$\thanks{ORCID: 0000-0002-9424-2339}
\\
$^{1}$Department of Astronomy and Space Sciences, Faculty of Science, Ege University, 35100, \.Izmir, Turkey\\
$^{2}$Department of Astronomy and Space Sciences, Graduate School of Natural and Applied Sciences, Ege University, 35100, \.Izmir, Turkey\\
}

\date{Accepted XXX. Received YYY; in original form ZZZ}

\pubyear{2015}

\begin{document}
\label{firstpage}
\pagerange{\pageref{firstpage}--\pageref{lastpage}}
\maketitle

\begin{abstract}
Asteroseismology provides a direct observational window into the structure and evolution of stars. While spectroscopic and photometric methods only offer information about the surface properties of stars, asteroseismology, through oscillation frequencies, provides comprehensive information about the deep stellar interior as well as the surface.
The scattering of effective temperature ($T_{\rm eff}$) determined from the spectrum and degeneracy in the Hertzsprung–Russel diagram poses challenges in developing a unique interior model for a single star. 
Although observational asteroseismic data partially lift this degeneracy, the best model that meets all asteroseismic constraints is not obtained. Most models reported in the literature typically address the large-separation ($\Delta\nu$) constraint between oscillation frequencies, 
which is a critical issue, especially in post–main sequence stars. Reference frequencies, influenced by helium ionisation zone–induced glitches in oscillation frequencies, are instrumental in refining models. Using the high metallicity derived from the colors of the Kepler Legacy star KIC 7747078, we obtain the mass of models ($M$) as $1.208$ $\rm M_{\sun}$ and $1.275$ $\rm M_{\sun}$ using the reference frequencies and individual frequencies as constraints, respectively.
By applying the $\chi^2$-method using these reference frequencies, $\Delta\nu$, and surface metallicity determined from the spectrum, we develop a unique star model with a mass of $1.171 \pm 0.019$ $\rm M_{\sun}$, a radius of $1.961 \pm 0.011$ $\rm R_{\sun}$, an effective temperature of 5993 K, an initial metallicity of 0.0121 and an age of $5.15 \pm 0.29$ Gyr. A significant advantage of this method is that $T_{\rm eff}$ emerges as an output, not a constraint. The mixed-mode oscillation frequencies of this model align well with the observations.
\end{abstract}




\section{Introduction}

Asteroseismology is one of the most effective scientific fields {that allows the study of} astronomical objects \citep{2004A&A...415..251D,2012ApJ...748L..10M,2018Natur.554...73G}. 
This field {examines the} oscillation frequencies, {in particular} those of solar-like oscillating stars, to help us better understand the internal structure and evolution of stars. Since such {oscillations have very low amplitude}, they {cannot be observed well} with ground-based telescopes. The CoRoT \citep{Baglin2006}, Kepler \citep{Borucki2010} and TESS \citep{Sullivan2015} space telescopes {have made it possible to observe the} oscillation frequencies of many main-sequence (MS) and evolved stars such as sub-giants (SGs) and red giants (RGs). 
Since these low-amplitude oscillations are excited by turbulent convection, such oscillations are observed in stars with convective envelopes. {These oscillations are either p-mode waves with acoustic properties or mixed modes. The latter behave like acoustic modes in the stellar envelope and like gravity modes in the core, containing important information about the structure of the stellar core.}  
Using the asteroseismic parameters derived from these observational oscillation frequencies, the fundamental properties of these stars can be accurately determined. These asteroseismic parameters are the frequency of maximum amplitude ($\nu_{\rm max}$), large separation ($\Delta\nu$) {and} small separation ($\delta\nu_{02}$) between oscillation frequencies and {the reference frequencies ($\nu_{\rm min}$)} {attributed to the} helium ionisation zone. These parameters and non-astoseismic parameters {(like effective temperature ($T_{\rm eff}$), visual magnitude ($V$) and parallax ($\pi$))} are used to determine the fundamental properties of the star, such as age ($t$), radius ($R$), mass ($M$), mean density ($\rho$) and gravity ($g$), by constructing interior models {using the Modules for Experiments in Stellar Astrophysics} ({\small MESA}) code \citep[]{Paxton2011, Paxton2013, Paxton2015, Paxton2018, Paxton2019, Jermyn2023}. {The parameters} $\rho$ and $g$ are related to $\langle\Delta\nu\rangle$ \citep{Ulrich1986} and $\nu_{\rm max}$ \citep{Brown1991}, respectively:
\begin{equation}
    {\langle\Delta\nu\rangle}\propto\sqrt\frac {M}{R^{3}}\propto\sqrt{\rho},
    \label{eq:largeseparation}
\end{equation}
\begin{equation}
    {\nu_{\rm max}}\propto\frac{\rm g}{\sqrt{T_{\rm eff}}}.
    \label{eq:frequencyofmaximum}
\end{equation}
%
These relations are very useful to obtain the so-called
conventional scaling relations \citep{Kjebed1995}. According to these relations mass and radius depend on $\langle\Delta\nu\rangle$, $\nu_{\rm max}$, and $T_{\rm eff}$:
\begin{equation}
\frac{M_{\rm sca}}{M_{\sun}}=\left( \frac{\nu_{\rm max}}{\nu_{\rm max\sun}}\right)^{3} 
\left( \frac{\langle\Delta\nu\rangle}{\langle{\Delta\nu_{\sun}}\rangle}\right)^{-4} 
\left( \frac{T_{\rm eff}}{T_{\rm eff \sun}}\right)^{1.5}
\label{eq:Msca}
\end{equation}
\begin{equation}
\frac{R_{\rm sca}}{R_{\sun}}=\left( \frac{\nu_{\rm max}}{\nu_{\rm max\sun}}\right)^{} 
\left( \frac{\langle\Delta\nu\rangle}{\langle{\Delta\nu_{\sun}}\rangle}\right)^{-2} 
\left( \frac{T_{\rm eff}}{T_{\rm eff \sun}}\right)^{0.5}.
\label{eq:Rsca}
\end{equation}
Parameters with the symbol $\sun$ represent the solar values. The accepted values are $\langle\Delta\nu_{\sun}\rangle=135.1\pm0.1$ $\mu$Hz, $\nu_{\rm max\sun}=3090\pm30$ $\mu$Hz, and $T_{\rm eff \sun}=5772\pm0.8$ K \citep{Huber2011,Prsa2016}. {However}, $\nu_{\rm max}$ {cannot} always be determined easily, and its uncertainty is not always low. For example, hot stars (e.g. Procyon) show double peaks in the power spectrum. Determining which of these peaks should be used in scaling relations is difficult. 
{To solve} this {problem}, \citet{Yildiz2019} derived alternative scaling relations using reference frequencies. If $\nu_{\rm min}$s can be determined from observational frequencies, $M$ and $R$ can be calculated accurately {using} alternative scaling relations. 

The radial order ($n$) and degree ($l$) of a mode determine the frequency of the mode ($\nu_{nl}$). While $\Delta \nu= \nu_{nl}-\nu_{n-1,l}$, {the term} $\delta \nu_{02}$ is the difference between $\nu_{n0}$ and $\nu_{n-1,2}$. {Here}, $\delta\nu_{\rm 02}$ is sensitive to the age of the MS stars \citep{Christensen1988}. 
The cavity for the oscillations with $l = 0$
is the whole star, including the centre. However, the turning point of the modes with $l=2$ is near the boundary of the nuclear core. 
The mean molecular weight in the
core increases as hydrogen is converted into helium. The increase
in the mean molecular weight leads to a decrease in the speed of sound.
The change in the core region directly affects the frequencies of modes with $l=0$ but not {those} of modes with $l=2$. Therefore, from {zero-age main-sequence (ZAMS) to thermal-age main-sequence (TAMS) stars}, the nuclear evolution in MS {stars} reduces $\delta\nu_{02}$.

Solar-like oscillations observed at different evolutionary stages of a star (MS, SG and RG) have {motivated} many studies on stellar structure and evolution. For example, the mixed modes observed in evolved stars {contain} information about the internal structure {of the star}. Using the observed mixed mode frequencies in SG and RG stars, {we can obtain} strong constraints on the convective core of the stars \citep{Deheuv2011,Aren2017,Montalban2013,CYO2023}. \citet{Noll2021} obtained strong constraints on the extension of the MS convective core using two g-dominant mixed mode frequencies of the star KIC 10273246. Furthermore, using $\Delta\nu$ and the g-dominant mixed mode, they estimated the core density with a precision of 1\% {to explain} why the high-value core overshooting differs from observations. Similarly, in solar-like oscillating MS stars, convective core constraints are determined using the ratio of small separation to large separation of the radial and dipole modes \citep{Silva2011,Tian2014,Deheuv2016}. {For evolved stars that show mixed modes}, the period spacing parameter ($\Delta\Pi$) determined from the $l=1$ frequencies of these stars is sensitive to the core region. This parameter can be used to distinguish between {the red clump and red giant branch} stars, which are difficult to distinguish in the Hertzsprung$-$Russel (HR) diagram \citep{Mosser2014,Bedding2011}. In addition, the rotational properties of stars are obtained {from} solar-like oscillations. {Rotation induces rotational splitting of non-radial oscillations,} and these splittings are used to obtain the rotation profile of the interior. In studies {on RGs and SGs}, rotational splitting was used to confirm that the core {of the star} rotates faster than the outer envelope \citep{Beck2012,Deheuv2012,Ahlborn2020}.

{It is possible to construct} more than one internal structure model {that matches} the position of the star in the HR diagram. For MS stars, these models can be distinguished from each other using seismic parameters such as $\delta\nu_{\rm 02}$. However, in evolved stars, {because of changes in the core, the sensitivity of the parameter to the core structure decreases}.

{
Since other seismic parameters such as $\Delta\nu$ and $\nu_{\rm max}$ are related to the general properties of the star, models that fit different values of $M$ and $R$ can be obtained. Yıldız et al. (2019) examined the discrepancies between the masses $M_{\rm sca}$ and radii $R_{\rm sca}$ obtained from scaling relations and those obtained from internal structure models (mass = $M_{\rm lit}$ and radius = $R_{\rm lit}$) of 83 stars based on asteroseismic data. Interestingly, in the ($\delta M/M=M_{\rm lit} - M_{\rm sca})/M_{\rm sca}$) versus ($\delta R/R=R_{\rm lit} - R_{\rm sca})/R_{\rm sca}$) graph, all stars clustered near the $\delta M/M=3 \delta R/R$ line. This result indicates that asteroseismic internal structure models well fit the observed $\Delta\nu$. However, mass differences can reach up to 20 $\%$ and radius differences up to 7 $\%$, indicating that most models reported in the literature lack uniqueness. We already know that equations 3 and 4 require correction \citep{2011ApJ...743..161W, 2016ApJ...822...15S, 2016MNRAS.462.1577Y, YildizOrtel2021}. For these reasons, we cannot determine a unique model of an evolved star using only $\Delta\nu$ and $\nu_{\rm max}$.

Another problem encountered when modeling stars is the difficulty in simultaneously fitting $\Delta\nu$ and individual frequencies to observed values. Typically, when a model's $\Delta\nu$ matches the observed value, its individual frequencies do not match, or vice versa. This discrepancy is often attributed to surface effects, leading to the acceptance of one solution over another, which complicates the discussions of model uniqueness.

Unknown chemical composition and age hinder the development of unique interior models. Even if we accurately determine $M$ and $R$, multiple similar models can emerge with different chemical compositions and ages. Research by \cite{2015MNRAS.448.3689Y} demonstrated how reference frequencies vary with chemical compositions in models generated by the ANK\.I evolution code  \citep{1965CaJPh..43.1497E, 2011MNRAS.412.2571Y}. This indicates that reference frequencies can be used  in obtaining a unique model.} 

In this study, we aim to determine the unique interior
model of KIC 7747078 using reference frequencies and asteroseismic (
$\delta\nu_{02}$, $\nu$, $\nu_{min}$, etc.) and non-asteroseismic ($T_{\rm eff}$, luminosity ($L$), $R$, etc.) observational parameters. Then, the model age and fundamental parameters of the star {can} be accurately determined.
However, many factors influence star evolution. {Among them, mass is the} best-known factor, which greatly influences star evolution.
{Similarly, the chemical composition also influences stellar evolution considerably.} Depending on the abundance of heavy elements, 
stellar evolution slows down or speeds up. For example, {among} two stars {with} the same mass {and} different metallicities, the one with a higher metallicity is expected to evolve more slowly. 
Therefore, in this study, models with various $Z_0$ and $Y_0$ values were constructed to understand the effect of chemical composition on the star.  We also test the effects of nuclear reaction rates and the chemical mixture of opacity tables for low-temperature regimes.

{The rest of this} paper is organised as follows: Section 2 presents the observational data of KIC 7747078. {Previously published works on} this star are presented in Section 3. Section 4 explains {the calculation of} the mass of the star and the model computation method. The model outputs are discussed in Section 5. Finally, Section 6 presents the conclusions of the study.



\section{Observational Data}
\begin{figure}
    \centering
   \includegraphics[width=\columnwidth]{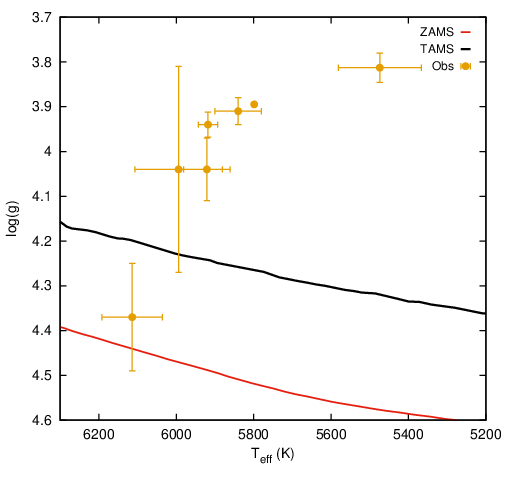}
    \caption{$\log{g}-T_{\rm eff}$  diagram. The filled yellow circles represent the values {reported} in the literature for KIC 7747078. {Zero-age main-sequence (solid red line) and thermal-age main-sequence (solid black line) are plotted using the models constructed by employing the solar composition \citep{Yildiz2015}.} }
    \label{fig:HRlocations}
\end{figure}

KIC 7747078 is a subgiant star and was observed by the Kepler space telescope. Its non-asteroseismic and asteroseismic observational data compiled from the literature {are summarised} in Tables \ref{tab:obs_nonasteroseismic} and \ref{tab:obs_asteroseismic}, respectively. 
{In Fig. \ref{fig:HRlocations}, the observational data of KIC 7747078 are plotted on the $\log{g} - T_{\rm eff}$ diagram.} The ZAMS and TAMS are represented by the red (bottom) and black (top) lines, respectively. 
The models that form these ZAMS and TAMS lines are constructed with the solar composition {derived} from the ANK\.I code \citep{Yildiz2015}. The filled yellow circles show the observational data for KIC 7747078 in Table \ref{tab:obs_nonasteroseismic}. 
\begin{table}
    \centering
    \caption{Effective temperature (${T_{\rm eff}}$), gravity ($\log{g}$) and metallicity ($[{\rm Fe}/{\rm H}]$) of KIC 7747078. {The references are given in the last column. MZ13 represents \citet{Molenda2013}}.}
    \begin{tabular}{llll}
        \hline
        $T_{\rm eff}$(K) & $\log{g}$(cgs) & $[{\rm Fe}/{\rm H}]$ & {\rm Reference}\\
        \hline
        $5840\pm60$ & $3.91\pm0.03$ & $-0.26\pm0.06$ & \citet{Bruntt2012}\\
        $6114\pm78$ & $4.37\pm0.12$ & $-0.11\pm0.06$ &  MZ13\\
        $5994\pm113$ & $4.04\pm0.23$ & $-0.19\pm0.23$ & MZ13\\
        $5918\pm25$ & $3.94\pm0.028$ & $-0.14\pm0.01$ & \citet{Brewer2016}\\
        $5921\pm60$ & $4.04\pm0.07$ & $-0.21\pm0.04$ & \citet{Furlan2018}\\
        $5798.709$ & $3.895$ & $-0.304$ & \citet{Ting2019} \\
        $5474.1^{+110.7}_{-103.5}$ & $3.813^{+0.032}_{-0.033}$ & $-0.231^{+0.155}_{-0.152}$ & \citet{Berger2020}\\
        \hline
    \end{tabular}
    \label{tab:obs_nonasteroseismic}
\end{table}
\begin{table}
    \centering
    \caption{Frequency parameters of KIC 7747078. Columns include the frequency of maximum amplitude (${\nu_{\rm max}}$), 
    large separation (${\Delta\nu}$) and references. {The maximum difference between ${\nu_{\rm max}}$ values is $\sim$5$\%$. }}
    \begin{tabular}{ccc}
        \hline
        ${\nu_{\rm max}}$(${\mu{\rm Hz}}$) & ${\Delta\nu}$(${\mu{\rm Hz}}$) & Reference\\
        \hline
        977 & 53.7 & \cite{Appo2012}\\
        936$\pm$32 & 53.9$\pm$0.3 & \citet{Chaplin2014}\\
        931$\pm$5 & 53.22$\pm$0.01 & \citet{LiY2020}\\
        \hline
    \end{tabular}
    \label{tab:obs_asteroseismic}
\end{table}
The visual magnitude ($V$) of the star is taken as 9.52 mag {from} the SIMBAD database. The distances ($d$) of the star determined from Gaia DR2 and EDR3 data are 180.1 and 179.958 pc, respectively \citep{Gaia2018,Gaia2021}. The values of ${T_{\rm eff}}$ (5474$-$6114 K), $log{\rm g}$ (3.81$-$4.37) and $[{\rm Fe}/{\rm H}]$ ($-0.30$ to $-0.11$ dex) are {reported in previous} studies \citep{Bruntt2012,Molenda2013,Brewer2016,Furlan2018,Ting2019,Berger2020}. 
{There is a very large scatter in the $\log{g}$ and $T_{\rm eff}$  values in particular. While $\log{g}$  can be calculated using equation 2, the scatter in $T_{\rm eff}$ raises concerns about its use as a direct constraint for developing interior models (Section 5).}

The observational asteroseismic parameters ${\nu_{\rm max}}$ and ${\Delta\nu}$ are {reported} in three {previous} studies \citep{Appo2012,Chaplin2014,LiY2020}. {The values of ${\nu_{\rm max}}$ and ${\Delta\nu}$ are} 931$-$977 and 53.2$-$53.9, respectively. Finally, the observed frequencies of the star are taken from \cite{Appo2012} and \citet{LiY2020}. Observational frequencies are {listed} in the first two columns in Table \ref{tab:allfrequencies}.
\section{Previous Model studies on KIC 7747078}
KIC 7747078 has left the {MS phase} and is in the SG phase. 
{This star has been widely studied.}
\citet{Chaplin2014} analysed more than 500 MS and SG stars, including KIC 7747078. They derived the fundamental parameters of KIC 7747078 from two different data sets using Kepler's asteroseismic data.
In the first method, they used $T_{\rm eff}$ = $5418\pm176$ K and $[{\rm Fe}/{\rm H}]$ = $-0.20\pm0.30$, {as} determined {by} the infrared flux method (IRFM), to estimate $M$, $R$ and $t$ {as} $1.04^{+0.12}_{-0.11}$ $\rm M_{\sun}$, $1.89\pm0.08$ $\rm R_{\sun}$ and $8.9^{+3.2}_{-2.3}$ Gyr{, respectively}.
In the second method, they used the spectral $T_{\rm eff}$ = $5840 \pm 84$ K and $[{\rm Fe}/{\rm H}]$= $0.26 \pm 0.09$ values obtained by \citet{Bruntt2012}. The values
of $M$, $R$ and $t$ determined {by the second method} are $1.10\pm0.06$ $\rm M_{\sun}$, $1.93 \pm 0.04$ $\rm R_{\sun}$ and $6.5 \pm 0.8$ Gyr, respectively.

On the other hand, \citet{Metcalfe2014} analysed 42 solar-like oscillating MS and SG \texttt{Kepler} targets using spectroscopic and asteroseismic data. They constructed interior models of the stars using the Asteroseismic Modelling Portal. {They obtained the following results for KIC 7747078:} $M=1.06\pm0.05$ ${\rm M_{\sun}}$, $R=1.889\pm0.023$ ${\rm R_{\sun}}$ and $t=6.26\pm0.92$ Gyr. They also estimated {the abundance of} heavy elements ($Z$), helium abundance ($Y_0$) and mixing length parameter ($\alpha$). The model input parameters are $Z = 0.0103 \pm 0.0019$, $Y_{\rm i} = 0.271 \pm 0.020$ and $\alpha=1.76 \pm 0.20$.

\citet{Yildiz2019} analysed the \textit{Kepler} and \textit{CoRoT} stars in the RG, SG and MS evolutionary phases {and} derived a new scaling relation from the interior
models. {In addition, they used the reference frequencies (${\nu_{\rm min0}}$ and ${\nu_{\rm min0}}$) to calculate the mass ($M_{\rm sis0}$ and ${M_{\rm sis1}}$) and radius ($R_{\rm sis0}$ and ${R_{\rm sis1}}$). The fundamental parameters they determined for this star are as follows: ${M_{\rm sca}} = 1.12 \pm 0.16$ ${\rm M_{\sun}}$, ${R_{\rm sca}} = 1.94 \pm 0.10$ ${\rm R_{\sun}}$, ${M_{\rm sis0}} = 1.10 \pm 0.02$ ${\rm M_{\sun}}$, ${M_{\rm sis1}} = 1.12 \pm 0.02$ ${\rm M_{\sun}}$, ${R_{\rm sis0}} = {R_{\rm sis1}} = 1.95 \pm 0.01$ ${\rm R_{\sun}}$. They also determined the age and initial metallicity as  ${t_{\rm sis}} = 4.00\pm0.20$ Gyr and ${Z_{\rm 0}} = 0.0128\pm0.0007$, respectively.}

Finally, \citet{LiT2020} modelled 36 Kepler SG stars and determined their ages. The mass, radius, luminosity and age estimates {for KIC 7747078} are $1.11\pm0.06$ ${\rm M_{\sun}}$, $1.92\pm0.03$ ${\rm R_{\sun}}$, $4.00\pm0.25$ ${\rm L_{\sun}}$ and $6.22\pm0.76$ Gyr, respectively. {The} results obtained by \cite{2021MNRAS.503.4529C} are similar to the results reported by \citet{LiT2020}.
\section{Methods}
\subsection{Computation of mass and radius}
\label{sec:MRcomput}
Stellar mass is one of the most influential parameters {for} stellar structure and evolution. Therefore, {it is crucial to determine the mass accurately}. Different methods can be used to determine the mass of solar-like oscillating stars.  For single stars, {despite some} uncertainty, stellar mass is calculated {by} the classical {method} based on parallax ($\pi$) (see below).  
To estimate the mass of a star using this method, we need distance ($d_{\pi}$), magnitude, $[{\rm Fe}/{\rm H}]$, $T_{\rm eff}$ and $\log{g}$ {data}. {The} Gaia mission {has provided data on} the distance values of most of the Kepler target stars. Since KIC 7747078, too, is observed by Gaia, the fundamental parameters can be calculated {using} the classical {method}. 

The first step in this method is to calculate the absolute magnitude ($M_{\textit{V}}$). $M_{\textit{V}}$ is obtained using $d_{\pi}$ and the apparent magnitude $V$. In the next step, the bolometric magnitude ($M_{\rm bol}$) is obtained from the absolute magnitude and bolometric correction tables \citep{Lejeune1998}. {Then, $L$ of KIC7747078 is calculated from $M_{\rm bol}$. Then, $R_{\pi}$ is estimated as {follows}:
\begin{equation}
    \frac{R_{\pi}}{R_{\sun}}=\left(\frac{L}{L_{\sun}}\right)^{0.5} \left(\frac{T_{\rm eff\sun}}{T_{\rm eff}}\right)^2 .
    \label{eq:lum_relation}
\end{equation}
Finally, using the $g$ value {determined} from the spectral or asteroseismic $\log{g}$ and the calculated $R_{\pi}$ value, we estimate the mass ($M_{\pi}$) as follows:
\begin{equation}
    \frac{M_{\pi}}{M_{\sun}}=\left(\frac{g}{g_{\sun}}\right)\left(\frac{R}{R_{\sun}}\right)^{2} .
    \label{eq:gravity}
\end{equation}
Many observational parameters {influence} the uncertainty {in} $M_{\pi}$. {The accurate estimation of} $M_{\pi}$ depends on {the accuracy of} observational $\log{g}$.

The asteroseismic parameter $\langle\Delta\nu\rangle$ {can be used} to test the accuracy of the estimated mass calculated by this method. {Toward} this end, we construct evolutionary models of different masses. $L$, $R$ and $T_{\rm eff}$ of the models are fitted to the observed values. For a given value of $R$, changing the $M$ {value} causes {a change in} the mean density. Models with different masses have different $\langle\Delta\nu\rangle$ values ($\langle\Delta\nu_{\rm mod}\rangle$). 
By comparing $\langle\Delta\nu\rangle$ and $\langle\Delta\nu_{\rm mod}\rangle$, we can determine {the model with the} more appropriate {mass}.

The compiled observational data of KIC7747078 are {listed} in Table 1. Different observational data yield different values for the fundamental parameters of the star. The evolutionary tracks of the models constructed using these fundamental parameters also differ from each other. The results of the constructed models are discussed in Section \ref{sec:RnD}.

The second method is based on the computation of $M_{\rm sca}$ and $R_{\rm sca}$ from equations (\ref{eq:Msca}) and (\ref{eq:Rsca}), respectively. 
KIC 7747078 is {an SG star}; however, equations (\ref{eq:Msca}) and (\ref{eq:Rsca}) are scaled to {an MS} star, namely the Sun.
Therefore, more general forms of equations (\ref{eq:Msca}) and (\ref{eq:Rsca}) are required for the evolved stars \citep{2011ApJ...743..161W, YildizOrtel2021, Yildiz2023}. In this method, we use $R_{\rm sca}$ {instead} of  $R_{\pi}$ in the classical method. Since the uncertainty in $M_{\rm sca}$ is much greater than {that} in $R_{\rm sca}$, we construct interior models for KIC 7747078 {with} the same radius but different masses. Then, we compute the adiabatic oscillation frequencies and obtain the $\langle\Delta\nu_{\rm mod}\rangle$ and $\nu_{\rm min}$s values of each model. {By comparing} the model and the observational values of $\langle\Delta\nu\rangle$ and $\nu_{\rm min}$s, we can find the suitable value of $M_{\rm mod}$ for a given value of $R_{\rm sca}$. For fine-tuning $\langle\Delta\nu\rangle$ and $\nu_{\rm min}$s simultaneously, we slightly change $R_{\rm mod}$ in some cases.
\subsection{Properties of the {\small {MESA}} evolution code}
\label{sec:MESA}
The interior models are constructed {using star and astero modules of the {\small {MESA}} (version 15140)} stellar evolution code
\citep[][]{Paxton2011, Paxton2013, Paxton2015, Paxton2018, Paxton2019, Jermyn2023}. 
{Based on} the calibration of the solar model, the {values of} $Y_{\sun}$ $Z_{\sun}$, and convective parameter ($\alpha_{\sun}$) values are obtained as 0.2745, 0.0172, and 1.814, respectively. 

The standard mixing length theory of \citet{Bohm1958} is used for convection. The interior models do not consider convective overshooting. The opacity for high $T$ values is determined from {\small OPAL} \citep{Iglesias1993, Iglesias1996}{, while that} for the low $T$ values {are taken} from {the tables in} \cite{Ferguson2005}. In addition, the Nuclear Astrophysics Compilation of Reaction Rates ({\small NACRE}) tables \citep[][]{Angulo1999} are used for nuclear reaction rates.
We also test different rates for the $^{14}$N(p,$\gamma$)$^{15}$O reaction. These rates are controlled with 
{\tt set\_rate\_n14pg = 'jina reaclib'} and {\tt set\_rate\_n14pg = 'NACRE'} in {\small MESA} (see Section 5).
Element diffusion is applied using {the method proposed by} Paquette et al. (1986). 
Atmospheric conditions affect the adiabatic oscillation frequencies {of the model}. The {option} \texttt{simple\_photosphere} in the {\small MESA} code {is used} for the stellar interior models. In this study, the pre-MS phase is included in the construction of the stellar interior models.

The adiabatic oscillation frequencies of the evolutionary model are calculated using the {\small ADIPLS} package \citep{Christensen2008}. In interior models, the surface layers of the star cannot be modelled exactly. Therefore, the \citet{Kjeldsen2008} surface-effect correction is {applied} to the model frequencies.

{
To compare the mixed mode frequencies of models and those obtained from observations, we use the astero module of {\small MESA}. This module includes a list of observational frequencies in its input file. Near-surface effects are computed using the "combined" option of \cite{2014A&A...568A.123B}.}
\subsection{Modelling method and uncertainty calculations}
\label{sec:modelmethod}
{Input parameters needed to construct an interior model are $M$, initial helium abundance ($Y_0$), initial metallicity ($Z_0$) and $\alpha$. 
In this work, we {uses the masses} $M_{\pi}$ and $M_{\rm sca}$ as the initial mass (Section \ref{sec:MRcomput}). $Z$ is obtained either from colours or from }the $[{\rm Fe}/{\rm H}]$ value determined from the spectra, and $\alpha$ is taken as the solar value. 
The last remaining input parameter $Y_0$ is used to fit the model to the position of the star in the HR diagram. $Y_0$ is changed so that the model parameters fit the observational {values of} $R$ and $L$. {Model} construction starts with $Y_0=Y_{\sun}$. If the model is not in agreement with the observations, a new model is computed with new $Y_0$. 
This process is repeated until a model compatible with the observation point in Fig. \ref{fig:evrim_modelC-D} is obtained. 

{In} the next step, model and observational seismic data (such as $\nu$, $\langle\Delta\nu\rangle$ and $\nu_{\rm min}$s) {are compared}. Suppose there is no agreement within the uncertainty; a new mass {value} is calculated to fit the observational $\langle\Delta\nu\rangle$ using the scaling relation. {This} process is repeated with a new value of $M$ and continues until an agreement is achieved within the uncertainty of $\langle\Delta\nu\rangle$.
The outputs of the models calibrated in this way are {listed} in Table \ref{tab:ABCD_modeldata}.

{Further, we obtain} unique models {using} the $\chi^2$-method. According to the surface metallicity ($Z_{\rm s}\sim 0.01$) of the star {determined} from its spectra, the initial metallicity is {approximately} 0.012. We obtain very dense grids in all the important model input parameters ($M$, $Z_0$, $Y_0$, and $\alpha$) at different ages (see Section 5.1.2). Then, we apply the normalised $\chi^2$ - method for the asteroseismic constraints. This method gives very small values of $\chi^2$, which are concentrated at a certain point in the multi-variable parameter space. {This aspect is considered a unique feature of the interior model if there is a single point at which $\chi^2$ is minimum in the multi-dimensional variable space. }

Uncertainties in {the values of} $M$, $R$, $\log{g}$, $L$ and $t$ computed in this study are estimated by Monte-Carlo simulations. The method is repeated on each synthetic data set, and the standard deviation of the distribution after 1000 iterations is taken as an estimate of the uncertainty.
The uncertainties in the outputs of the models are also {listed} in Table \ref{tab:ABCD_modeldata}.
\begin{figure}
	\includegraphics[width=\columnwidth]{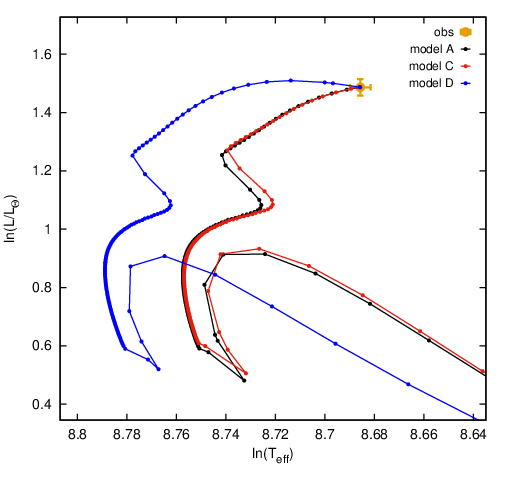}
    \caption{HR diagram. The diagram shows the evolutionary tracks of Models A, C and D. Although the models fit the same observation point, they have different evolutionary tracks. Models A, C and D are represented by solid circles in black, red and blue, respectively. The filled yellow circle indicates the observation point on the HR diagram for KIC 7747078.}
    \label{fig:evrim_modelC-D}
\end{figure}
\section{Results and Discussions}
\label{sec:RnD}
The small separation is a strong age indicator for the solar-like oscillating stars.  
{The value of} $\delta\nu_{02}$ is maximum around the ZAMS, {while} it is approximately 5-6 ${\mu}$Hz around the TAMS.
In evolved stars, the sensitivity of the small separation to the core structure decreases as the internal structure of the star changes. 
Therefore, strong constraints and precise observational data are needed to construct a unique model for evolved stars. These observational data are used in the methods described in Section \ref{sec:MRcomput} to estimate the fundamental properties of the star, such as $M$ and $R$. 

The fact that the luminosity of KIC 7747078 is close to its maximum ($L_{\rm max}$) after TAMS {is advantageous for modelling}. Luminosity in the HR diagram ($L\approx L_{\rm max}$) {remains} almost constant with age as it is determined by $M$, $Z_0$, and $Y_0$. {When} multiple parameters {are involved}, the ineffectiveness of some parameters brings the solution closer to a unique solution. For such stars, for example, we can construct models whose $\alpha$ and age {values} differ {while} the remaining features are the same (see Models A and D in Table \ref{tab:ABCD_modeldata}), and we can examine their asteroseismic properties.

\begin{table*}
    \centering
    \caption{Fundamental parameters of the models. The columns {present the} model name, mass ($M$), radius ($R$), effective temperature ($T_{\rm eff}$), metallicity ($Z_0$), helium abundance ($Y_0$), convective parameter ($\alpha$), age ($t_9$), reference frequencies ($\nu_{\rm min0}$ and $\nu_{\rm min1}$), large separation (${\Delta\nu}$), frequency of the maximum amplitude ($\nu_{\rm max}$) and the mean of the difference between the model and observed frequencies ($\overline{d\nu}$). The last row of the table lists the observed $\nu_{\rm min0}$ and $\nu_{\rm min1}$ values sourced from \citet{Yildiz2019} and $\Delta\nu$ and $\nu_{\rm max}$ sourced from \citet{LiY2020}.} 
    \begin{tabular}{ccccccccccccc}
        \hline   Model&$M$&$R$&$T_{\rm eff}$&$Z_{0}$&$Y_{0}$& ${\alpha}$ &$t_9$ &${\nu_{\rm min0}}$& ${\nu_{\rm min1}}$ & $\langle\Delta\nu\rangle$& ${\nu_{\rm max}}$ & $\overline{d\nu}$ \\[1.2pt]
        &(${M_{\sun}}$)&(${R_{\sun}}$)&(K)& & & &(Gyr)&(${\mu {\rm Hz}}$)&(${\mu {\rm Hz}}$)&(${\mu {\rm Hz}}$)&(${\mu {\rm Hz}}$) & (${\mu {\rm Hz}}$) \\[1.2pt]
         \hline
        A & 1.208 & 2.00 & 5918 & 0.0161 & 0.2702 & 1.814 & 5.24 & 1038.21 & 794.89 & 53.22 & 921.60 & 10.57\\[2.0pt]
          & $\pm$0.021 & $\pm$0.01 & $\pm$25 & $\pm$0.0003 & $\pm$0.0040 & --- & $\pm$1.06 & $\pm$12.13 & $\pm$9.29 & $\pm$0.31 & $\pm$19.32 & --- \\[2.0pt]
        B & 1.308 & 2.07 & 5918 & 0.0200 & 0.2584 & 1.814 & 4.77 & 1068.24 & 802.60 & 52.57 & 931.54 & 21.02 \\[2.0pt]
          & $\pm$0.022 & $\pm$0.01 & $\pm$25 & $\pm$0.0003 & $\pm$0.0040 & --- & $\pm$0.96 & $\pm$12.54 & $\pm$9.42 & $\pm$0.31 & $\pm$18.94 & --- \\[2.0pt]
        C & 1.213 & 2.00 & 5918 & 0.0172 & 0.2763 & 1.814 & 5.11 & 1049.34 & 813.81 & 53.33 & 925.41 &  8.87 \\[2.0pt]
          & $\pm$0.020 & $\pm$0.01 & $\pm$25 & $\pm$0.0003 & $\pm$0.0050 & --- & $\pm$1.02 & $\pm$12.37 & $\pm$9.59 & $\pm$0.31 & $\pm$19.23 & --- \\[2.0pt]
        D & 1.208 & 2.00 & 5918 & 0.0161 & 0.2702 & 2.454 & 5.42 & 1004.01 & 783.07 & 53.27 & 921.60 & 4.16 \\[2.0pt]
          & $\pm$0.021 & $\pm$0.01 & $\pm$25 & $\pm$0.0003 & $\pm$0.0040 & --- & $\pm$1.10 & $\pm$8.28 & $\pm$6.46 & $\pm$0.22 & $\pm$19.84 & --- \\[2.0pt]
        E & 1.275 & 2.02 & 6160 & 0.0220 & 0.3060 & 2.454 & 4.04 & 1093.25 & 831.90 & 53.61 & 934.63 & 0.10 \\[2.0pt]
          & $\pm$0.021 & $\pm$0.01 & --- & $\pm$0.0004 & $\pm$0.0050 & --- & $\pm$0.84 & $\pm$10.93 & $\pm$8.32 & $\pm$0.27 & $\pm$20.22 & --- \\[2.0pt]
        F & 1.279 & 2.03 & 6094 & 0.0267 & 0.3200 & 2.454 & 4.02 & 1118.59 & 833.43 & 53.42 & 933.36 & 2.78 \\[2.0pt]
          & $\pm$0.022 & $\pm$0.01 & --- & $\pm$0.0005 & $\pm$0.0050 & --- & $\pm$0.82 & $\pm$11.23 & $\pm$8.37 & $\pm$0.27 & $\pm$19.65 & ---- \\[2.0pt]
       obs & --- & --- & --- & --- & --- & --- & --- & 1039.3 & 792.5 & 53.22 & 931 & ---\\[2.0pt]
           & --- & --- & --- & --- & --- & --- & --- & $\pm$10.4 & $\pm$7.9 & $\pm$0.01 & $\pm$5 & ---\\[2.0pt]
        \hline
    \end{tabular}
    \label{tab:ABCD_modeldata}
\end{table*}
\begin{table*}
    \centering
    \caption{Observed oscillation frequencies of the $l=0$ mode{, as reported} in the literature for KIC 7747078, and frequencies of the six models constructed in this study.}
    \begin{tabular}{cccccccc}
         \hline
         \cite{Appo2012} & \citet{LiY2020} & Model A & Model B & Model C & Model D & Model E & Model F \\[1.2pt]
         \hline
          559.634$\pm$----- &  559.83$\pm$0.101 &  551.311 &  544.715 &  552.393 &  556.462 &  559.099 &  557.222 \\[2.0pt]
          612.376$\pm$1.272 &  611.49$\pm$1.256 &  602.264 &  595.138 &  603.429 &  608.738 &  611.027 &  609.097 \\[2.0pt]
          664.495$\pm$0.201 &  664.40$\pm$0.307 &  654.707 &  646.486 &  655.948 &  661.831 &  663.703 &  661.649 \\[2.0pt]
          717.500$\pm$0.124 &  717.56$\pm$0.109 &  707.661 &  698.887 &  708.977 &  714.667 &  717.466 &  715.232 \\[2.0pt]
          769.836$\pm$0.183 &  770.02$\pm$0.065 &  759.983 &  751.102 &  761.573 &  766.703 &  770.674 &  768.350 \\[2.0pt]
          822.395$\pm$0.181 &  822.56$\pm$0.058 &  812.171 &  802.596 &  813.806 &  819.100 &  823.078 &  820.574 \\[2.0pt]
          876.056$\pm$0.124 &  876.06$\pm$0.054 &  865.246 &  854.825 &  866.866 &  872.405 &  876.145 &  873.333 \\[2.0pt]
          929.999$\pm$0.099 &  930.20$\pm$0.054 &  919.229 &  907.920 &  921.009 &  926.015 &  930.188 &  927.110 \\[2.0pt]
          984.210$\pm$0.108 &  984.28$\pm$0.064 &  973.263 &  961.439 &  975.123 &  979.528 &  984.633 &  981.307 \\[2.0pt]
         1037.893$\pm$0.212 & 1038.20$\pm$0.262 & 1027.193 & 1014.949 & 1029.235 & 1033.187 & 1039.014 & 1035.492 \\[2.0pt]
         1092.389$\pm$0.235 & 1092.89$\pm$0.256 & 1081.241 & 1068.240 & 1083.399 & 1087.195 & 1093.255 & 1089.554 \\[2.0pt]
         1146.166$\pm$0.582 & 1147.35$\pm$0.624 & 1135.465 & 1121.881 & 1137.644 & 1141.435 & 1147.695 & 1143.719 \\[2.0pt]
         1201.562$\pm$0.699 & 1202.24$\pm$1.209 & 1189.953 & 1175.592 & 1192.310 & 1195.738 & 1202.440 & 1198.250 \\[2.0pt]
         \hline
    \end{tabular}
    \label{tab:allfrequencies}
\end{table*}

\subsection{Interior models with high $Z$}
{The $T_{\rm eff}$ and $Z$ values can also be obtained from the ($B-V$) and ($V-K$) colours. Using the colour tables of \cite{Lejeune1998}, we estimate $T_{\rm eff}$ to be about 6160 K and $Z$ to be 0.0220. These values are significantly higher than those obtained from spectral analysis (see Table \ref{tab:obs_nonasteroseismic}).}

We first construct models fitted to the spectral and asteroseismic data in Tables \ref{tab:obs_nonasteroseismic} and \ref{tab:obs_asteroseismic}, respectively. Table \ref{tab:ABCD_modeldata} {lists} the results of all the calibrated models with the JINA reaction rate (Cyburt et al. 2010) for $^{14}$N(p,$\gamma$)$^{15}$O ({\tt set\_rate\_n14pg = 'jina reaclib'}). Although the models fit the parameters ($L$, $R$, $T_{\rm eff}$, etc.) determined from the observational data, models may have different $M$, $t$, $\alpha$, $\nu$ and $\nu_{\rm min}$s. Fig. \ref{fig:evrim_modelC-D} shows the evolutionary tracks of three models (Models A, C and D in Tables \ref{tab:ABCD_modeldata} and \ref{tab:allfrequencies}) with different $M$ and $\alpha$ values.  
$R$ and $T_{\rm eff}$ of these models are 2.00 $\rm R_{\sun}$ and 5918 K, respectively.
Despite the significant differences in their evolutionary tracks, the input parameters of models A and D are identical, except for their $\alpha$s. Model A has $\alpha_{\sun}$ {while the} 
$\alpha$ {value} of model D is greater than $\alpha_{\sun}$:  $\alpha=2.454$. The $T_{\rm eff}$ difference between models D and A is {approximately} 180 K during the MS phase but zero at the observed position of the star in the HR diagram. Since Model D is slightly more evolved than Model A, its age is slightly higher than that of Model A. Models A and D have the same mean density, and therefore, they must have identical $\Delta\nu$ values (Fig. \ref{fig:Dnu-nu_modelC-D}). We need to know if there is any difference between the asteroseismic properties of these two models (see below).

Models A and C have {equal} $\alpha$ 
{values} and slightly different masses. 
Model C has a slightly higher mass, and its age is slightly less than that of Model A. Additionally, {the $Z_0$ values} of {the two models differ by} around 0.001. It is not easy to {determine} which of these three models, which intersect at the same point on the HR diagram, represents KIC 7747078. Asteroseismic analysis can reveal which of these models best represents the star.
\begin{figure}
    \centering
    \includegraphics[width=\columnwidth]{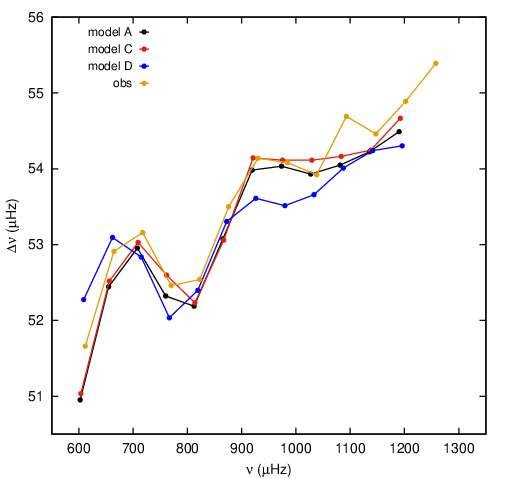}
    \caption{$\Delta\nu$ is plotted with respect to  $\nu$. The filled yellow circle indicates the observational frequencies of KIC 7747078.  The filled black, red and blue circles represent the adiabatic oscillation frequencies of Models A, C and D, respectively.}
    \label{fig:Dnu-nu_modelC-D}
\end{figure}
\begin{figure}
    \centering
   \includegraphics[width=\columnwidth]{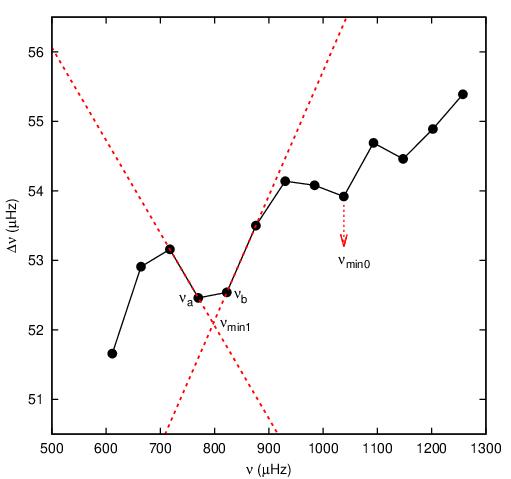}
    \caption{Graph of $\Delta\nu$ versus $\nu$ for {determining} the reference frequencies $\nu_{\rm min1}$ \textbf{and ${\nu_{\rm min0}}$}. The {reference frequency} $\nu_{\rm min1}$ is between $\nu_{\rm a}$ and $\nu_{\rm b}$. We plot two lines from the neighbourhood intervals. The intersection of these two lines gives us $\nu_{\rm min1}$. 
    ${\nu_{\rm min0}}$ is around an observed frequency. 
    The $\nu_{\rm min1}$ and $\nu_{\rm min0}$ values of the star are found to be 792.5 and 1039.3 ${\mu}$Hz, respectively.}
    \label{fig:find_numin}
\end{figure}
\begin{figure}
\centering
\includegraphics[width=\columnwidth]{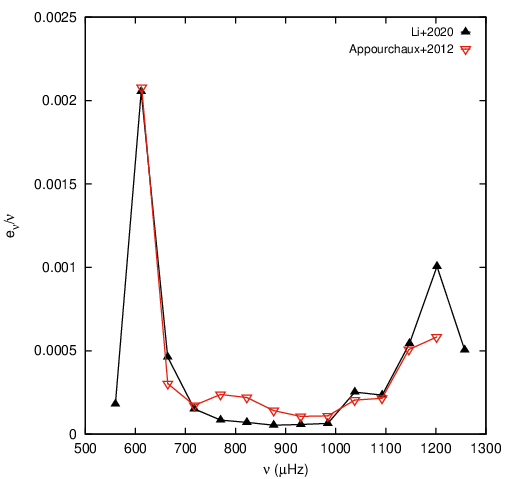} 
\includegraphics[width=\columnwidth]{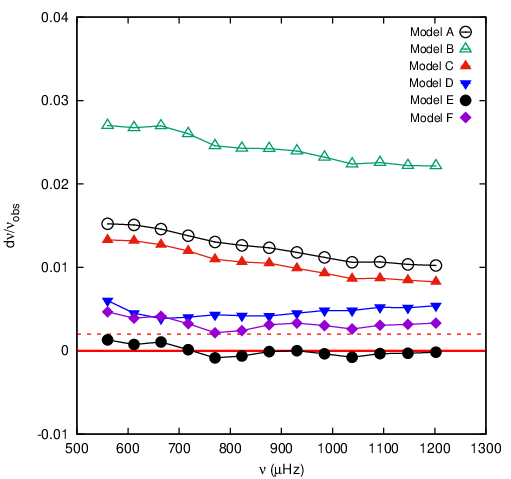} 
 \caption{The upper panel shows the uncertainty ($\rm e_{\nu}$) of the observation frequencies of KIC 7727078 that is plotted with respect to $\nu_{\rm obs}$. The filled black triangles and the hollow red triangles represent the frequencies as reported by \citet{LiY2020} and \citet{Appo2012}, respectively. The lower panel plots the fractional difference between the observational and model frequencies with respect to $\nu_{\rm obs}$. The graph shows the results {obtained by} six models. Model E provides the best agreement with the observational frequencies. 
 }
\label{fig:obs_freq_error_rate}
\label{fig:frequencies_ratio}
\end{figure}
\begin{figure}
    \centering
    \includegraphics[width=\columnwidth]{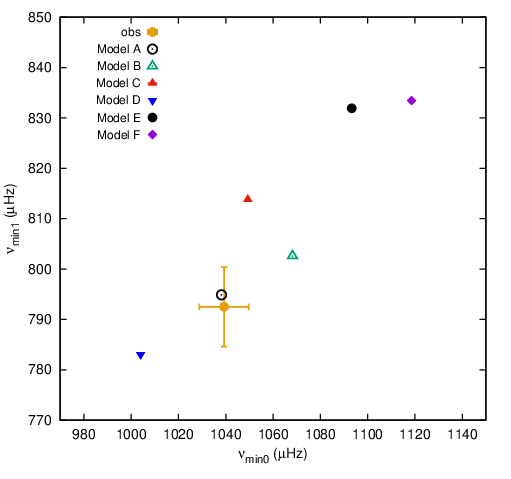}
    \caption{{The plot of} $\nu_{\rm min0}$ against $\nu_{\rm min1}$. The graph shows the reference frequencies of the six models. {The solid yellow circle with error bars represents the observational reference frequencies.} The observational values of $\nu_{\rm min1}$ and $\nu_{\rm min0}$ are 792.5 and 1039.3 $\mu$Hz, respectively. Model A gives a common solution for $\nu_{\rm min1}$ and $\nu_{\rm min0}$.}
    \label{fig:numin0-numin1}
\end{figure}

\subsubsection{Distinguishing models with $\nu_{\rm min}s$ and individual frequencies}
Although the model input parameters in Table \ref{tab:ABCD_modeldata} are very close to each other, it has sufficient coverage considering the possible $Z_0$ - $Y_0$ relationship and asteroseismic data. 
Since $\langle\Delta\nu\rangle$ and $\nu_{\rm max}$ are related to the general properties of the star (equations (\ref{eq:largeseparation}) and (\ref{eq:frequencyofmaximum})), they are not very suitable for distinguishing between models. Models A and D, for example, have the same mass and radius, and therefore their ${\Delta\nu}$ and $\nu_{\rm max}$ are very close to each other. 
For such cases, we also compare the individual oscillation frequencies of models with the observed frequencies. 
Individual oscillation frequencies of the models depend on the input parameters and surface effect. The difference between {the mean frequency of} Model A and the observed frequencies ($\overline{d\nu}=\nu_{n0,\rm obs}-\nu_{n0,\rm mod}$) is {approximately} 11 $\mu$Hz. This difference is very small compared to the individual frequencies and may be due to {surface} effects. Therefore, the use of individual oscillation frequencies alone may be {insufficient} to distinguish the models. {In such cases, different asteroseismic parameters, such as reference frequencies, are needed to distinguish models because $\nu_{\rm min}$s of Models A and D are significantly different.}

We investigate whether the models can be distinguished in terms of the reference frequencies introduced by \citet{Yildiz2014}. These are the {frequencies} corresponding to the minima in the $\Delta\nu$-$\nu$ graph due to the glitches induced by the second helium ionization zone on the oscillation frequencies.  These reference frequencies can be detected at the evolutionary stages MS, SG and RG. With evolution, the structure of the minima in the $\Delta\nu$-$\nu$ graph changes and shifts towards lower frequencies. $\nu_{\rm min1}$ is in general seen in the oscillation frequencies of all the solar-like oscillators. However, $\nu_{\rm min2}$ is available for the relatively hot oscillators, $\nu_{\rm min0}$ is more prominent in particular for the cooler oscillators. $\nu_{\rm min1}$ and $\nu_{\rm min0}$ of KIC 7747078 are shown in Fig. \ref{fig:find_numin}.

To find any $\nu_{\rm min}$, we first determine the frequency interval of a certain minimum in the graph of $\Delta\nu$ versus $\nu$. Two lines are drawn for the neighbourhood intervals. The intersection of the two lines gives the reference frequency $\nu_{\rm min}$. The method {to determine} $\nu_{\rm min}$ is shown in Fig. \ref{fig:find_numin}.
The value of $\nu_{\rm min}$ depends on the fundamental parameters of the star, such as $M$, $R$ and $T_{\rm eff}$.  
In the $\Delta\nu-\nu$ graph of KIC 7747078, the reference frequencies $\nu_{\rm min1}$ and $\nu_{\rm min0}$ correspond to the minima at low and high frequencies, {i.e.} 792.5 and 1039.3 ${\mu}Hz$, respectively. 

{When determining $\nu_{\rm min0}$, particular care is needed owing to the shallowness of the trough, which can make the determination of observational $\nu_{\rm min0}$ challenging (Fig. \ref{fig:find_numin}). Therefore, it is necessary to track how the minima change with evolution using grids. For KIC 7747078, an evolved star, mixed modes obscure clear minima in the $\Delta\nu$ - $\nu$ plot for modes with $l > 0$. Therefore, we use reference frequencies from modes with $l = 0$ to find the best model. 
In addition to these asteroseismic parameters, we use individual oscillation frequencies as constraints for interior models. In this case, we use the parameters of the best model obtained using Kjeldsen et al. (2008) method for the near-surface effects as the initial parameter and construct a new model (Model AN in Table 5) using astero modules with the "combined" option (Ball \& Gizon 2014). 
This model effectively fulfills all observational constraints, including mixed mode frequencies (Section \ref{sec:mixmode_astero}).}

{Asteroseismic methods generally provide more accurate results for $\log{g}$ compared to spectroscopic methods \citep{2013ARA&A..51..353C}. Therefore, asteroseismic $\log{g}$ is used as a benchmark when selecting spectral data.}
The spectral $\log{g}$ {reported} by \citet{Bruntt2012} is in good agreement with the seismic $\log{g}$ given by \citet{Yildiz2019} and \citet{LiY2020} 
 ($\log{g}_{\rm seis} = 3.91$).
 $T_{\rm eff}$ is 5840 K, as determined by \citet{Bruntt2012} from spectral analysis. First, this value of $T_{\rm eff}$ is used as one of the constraints for the interior models of KIC 7747078. {However,} for this {value}, neither the individual oscillation frequencies nor $\nu_{\rm min}$ {values agree} with the observational values. 

 The $\log{g}$ {value reported by} \citet{Brewer2016} is close to the $\log{g_{\rm seis}}$ {value}. The researchers {reported} $T_{\rm eff}$ to be 5918 K. The $\nu_{\rm min1}$ and $\nu_{\rm min0}$ {values} of the model constructed under constraints of this spectral data are 794.89 and 1038.21 $\mu$Hz, respectively. They are in close agreement with the observed $\nu_{\rm min}$ {values} (Model A in Table \ref{tab:ABCD_modeldata}).
However, there {exists} a mean difference of $11$ $\mu$Hz between the observational and model individual frequencies.


We can try to eliminate this difference. To increase the model frequencies by $11$ $\mu$Hz, we keep $R$ constant and estimate $M=1.213$  M$_{\sun}$ from the basic relationship between the stellar mean density and oscillation frequency of a certain mode. A model with this mass (Model C) has $\langle\Delta\nu\rangle$ =53.33 ${\mu}$Hz{; however, this value} should be around 53.67 ${\mu}$Hz according to the asymptotic relation. This {discrepancy} reveals that the asymptotic relation works slightly differently than our expectation. Although these values are very close, the difference is {considerable} if we compare the model results with the observational values of \cite{LiY2020}.   
When we compare Models C {and} A, {we see that} although the individual frequencies are improved {slightly}, the difference between the observational and model reference frequencies increases.

In Model A, both minima are in very good agreement with the observational data. However, there is a difference of ~11 $\mu$Hz between the observational and model oscillation frequencies. To eliminate this difference, the convective mixing length parameter $\alpha$ is changed. In Model D, the input parameters $M$, $Z_0$ and $Y_0$ are taken as the same as {those} in Model A. Only the parameter $\alpha$ is taken as 2.454 instead of $\alpha_{\sun}$.
Fig. \ref{fig:evrim_modelC-D} shows the evolutionary tracks of Model A (filled black circle) and Model D (filled blue circle). Increasing $\alpha$ increases the effective temperature of the star, and the evolutionary track {of the star} shifts to the left in the HR diagram. 
The effect of $\alpha$ on the individual oscillation frequencies {needs to be determined.} The {mean} difference between the model and observational individual frequencies decreases to 4 $\mu$Hz. 
The good agreement obtained for $\nu_{\rm min1}$ and $\nu_{\rm min0}$ in Model A {is not} achieved in Model D. The difference between observational and model $\nu_{\rm min1}$ becomes 10 $\mu$Hz. The difference for $\nu_{\rm min0}$ is 30 $\mu$Hz.

The $\nu_{\rm max}$ {values} of Models A and C are 6 $-$ 9 $\mu$Hz less than the observed $\nu_{\rm max}$ value  (Li et al. 2020a). {To realise} the agreement between the observed and model $\nu_{\rm max}$, we construct Model B with 1.308 M$_{\sun}$ and 2.07 R$_{\sun}$. The $\langle\Delta\nu\rangle$ of Model B is $52.57$ $\mu$Hz, {which is} less than the observed value. The mean difference between the observed and Model B frequencies is {approximately} 21 $\mu$Hz. This is the greatest difference {in} $\overline{d\nu}$ among the whole interior models {listed} in Table 3. This implies that {it is not easy to achieve} the simultaneous fit of $\langle\Delta\nu\rangle$,  $\nu$s and $\nu_{\rm max}$ of a model to the observed values. This result probably indicates that we need to improve the scaling relation for $\nu_{\rm max}$. 
However, $\nu_{\rm min1}$ and $\nu_{\rm min0}$ of Model B shift towards higher frequency values,
{reaching} 802.60 and 1068.24 $\mu$Hz, respectively.

\subsubsection{Effect of $T_{\rm eff}$ on individual frequencies}
%

Using the $\alpha$ {value} of Model D, we construct a new model with $T_{\rm eff}=6160$ K and $Z=0.0220$. {This model} is {labelled as} Model E in Table \ref{tab:ABCD_modeldata}. Its $\overline{d\nu}$ has the smallest value (0.1 $\mu$Hz) among the models. If we consider only the individual oscillation frequencies, Model E {is} the best model for KIC 7747078. However, $\nu_{\rm min1}$ and $\nu_{\rm min0}$ of Model E are significantly greater than the observed values. The difference between the model and observed values of $\nu_{\rm min}$s is about $\langle\Delta\nu\rangle$. 

We construct Model F with slightly modified parameters of Model E. Its $T_{\rm eff}$ and $Z_0$ are 6094 K and 0.0267, respectively. Both the reference and the individual frequencies of Model F are greater than those of Model E.

Table \ref{tab:allfrequencies} {lists the} observational and model frequencies for $l=0$. The first two columns are the observational frequencies given in the literature. The mean difference between the observational individual oscillation frequencies and that of Model E is {approximately} 0.1 $\mu$Hz. The top panel of Fig. \ref{fig:obs_freq_error_rate} shows the uncertainty of the oscillation frequencies ($e_\nu$) given in \citet{Appo2012} and \citet{LiY2020}. The uncertainty is large around $\nu = 600$ $\mu$Hz. If this datum is ignored, the fractional uncertainty ($\overline{d\nu}/\nu$) {decreases} to 0.001. Model frequencies are also analysed {by considering} these observational frequencies. The bottom panel of Fig. \ref{fig:frequencies_ratio} shows the fractional difference between \citet{LiY2020} and model frequencies. The dashed red line represents the upper limit of the uncertainty. Accordingly, Model B has the highest difference, while Model E is below the dashed line, giving a distribution around 0. 

Fig. \ref{fig:numin0-numin1} shows the reference frequencies plotted against each other. 
The observational position of KIC 7747078 ($\nu_{\rm min1}=792.5$ $\mu$Hz and $\nu_{\rm min0}=1039.3$ $\mu$Hz) is {plotted along} with error bars. The model closest to the observed position of the star in Fig. \ref{fig:numin0-numin1} is Model A. 
\begin{table*}
    \centering
    \caption{Basic properties of the models constructed using the $\chi^2-$ method. 
    {The columns are ordered as follows: model name, mass ($M$), radius ($R$), effective temperature ($T_{\rm eff}$), initial metallicity ($Z_0$), initial helium abundance ($Y_0$), convective parameter ($\alpha$), age ($t_9$), reference frequencies ($\nu_{\rm min0}$ and $\nu_{\rm min1}$), large separation (${\Delta\nu}$), frequency of maximum amplitude ($\nu_{\rm max}$), surface metallicity ($Z_s$), $\chi^2_{\rm seis}$, mixture of metallicity for low-opacity temperature \citep{Asplund2009, GS98} and module.}
    Models N1 $-$ N7 are derived using the NACRE reaction rate for $^{14}$N(p,$\gamma$)$^{15}$O. Model J is obtained with the JINA rate for that reaction. 
    {Model AN in the last row is constructed using the astero module.}
    {The} constraint for $\langle\Delta\nu\rangle$ is 53.4 $\mu$Hz {for Models N1 $-$ N4, and 53.22 $\mu$Hz for Models N5 $-$ N7 and J.
    Our favourite models are N1 and N6 {with} 53.4 and 53.22 $\mu$Hz, respectively. The basic parameters of the models are quite similar, {implying} that our method provides a unique model for KIC 7747078.} } 
    \begin{tabular}{cccccccccccccccc}
        \hline
        Model&$M$&$R$&$T_{\rm eff}$&$Z_{0}$ &$Y_{0}$& $\alpha$ &$t$&$\nu_{\rm min0}$& $\nu_{\rm min1}$& $\Delta\nu$& $\nu_{\rm max}$ & $Z_{s}$ & $\chi^2_{\rm seis}$ & mixture & module\\[1.2pt]
         &(M$_{\sun}$)&(R$_{\sun}$)&(K)& & &  & (Gyr)&($\mu {\rm Hz}$)&  ($\mu {\rm Hz}$)& ($\mu {\rm Hz}$)& ($\mu {\rm Hz}$) &  & & & \\[1.2pt]
         \hline
         N1 & 1.17 & 1.975 & 5977 & 0.0121 & 0.2689 & 1.8311 & 5.17 & 1037.4 & 791.2 & 53.40 & 910.76 & 0.0103 & 0.02 & gs98 & star\\[2.0pt]
         N2 & 1.17 & 1.975 & 5993 & 0.0120 & 0.2711 & 1.8311 & 5.08 & 1046.3 & 794.7 & 53.40 & 909.82 & 0.0102 & 0.18 & gs98 & star\\[2.0pt]
         N3 & 1.17 & 1.975 & 5996 & 0.0120 & 0.2711 & 1.8511 & 5.09 & 1050.7 & 792.6 & 53.40 &  909.71 & 0.0103 & 0.40& gs98 & star\\[2.0pt]
         N4 & 1.17 & 1.970 & 6048 & 0.0120 & 0.2711 & 1.8311 & 5.06 & 1040.6 & 799.1 & 53.51 & 909.78 & 0.0097 & 0.34& a09 & star\\[2.0pt]
         \hline
         N5 & 1.17 & 1.979 & 5988 & 0.0120 & 0.2711 & 1.8311 & 5.08 & 1032.1 & 790.4 & 53.22 & 905.97 & 0.0102 & 0.18& gs98 & star\\[2.0pt]
         N6 & 1.18 & 1.985 & 6005 & 0.0120 & 0.2687 & 1.8311 & 5.00 & 1039.4 & 796.2 & 53.22 & 907.47 & 0.0102 & 0.07& gs98 & star\\[2.0pt]
         N7 & 1.17 & 1.979 & 5990 & 0.0120 & 0.2711 & 1.8511 & 5.09 & 1033.4 & 788.7 & 53.22 & 905.85 & 0.0103 & 0.18 & gs98 & star\\[2.0pt]
         \hline
         J  & 1.16 & 1.973 & 6011 & 0.0120 & 0.2711 & 1.8511 & 5.26 & 1043.0 & 789.1 & 53.21 & 901.94 & 0.0100 & 0.10& gs98 & star\\[2.0pt]
        \hline
        AN & 1.17 & 1.961 & 5993 & 0.0121 & 0.2689 & 1.8311 & 5.15 & 1037.8 & 809.4 & 53.55 & 927.04 & 0.0103 & 0.34& gs98 & astero\\[2.0pt]
        \hline
    \end{tabular}
    \label{tab:NACRE_JINA}
\end{table*}

{The} $\langle\Delta\nu\rangle$ of all models {(Table \ref{tab:ABCD_modeldata})} are compatible with the observational $\langle\Delta\nu\rangle$. The $\nu_{\rm min1}$ and $\nu_{\rm min0}$ of Model A agree very well with the observational $\nu_{\rm min1}$ and $\nu_{\rm min0}$, while the individual observational frequencies of Model E agree very well with the observational frequencies.  The reference frequencies of Model E exceed the observational {values} by approximately $\langle\Delta\nu\rangle$ (54 and 39 $\mu$Hz). The difference between Model A and observational individual frequencies is 11 $\mu$Hz. If we {consider} only the asteroseismic data, Model A is much more consistent with observations because its $\overline{d\nu}<< \langle\Delta\nu\rangle$. The difference between individual frequencies is related to the surface effect. However, the {shortcoming} of Model A is its $Z_{\rm s}=0.0139$ ($[{\rm Fe}/{\rm H}] = +0.016$) {which} is significantly greater than the observed values {listed} in Table \ref{tab:obs_nonasteroseismic}.

The asteroseismic $\langle\Delta\nu\rangle$ parameter can be determined from the power spectrum. The most sensitive value for KIC 7747078 is 53.22 $\mu$Hz {reported} by Li et al. In cases where individual frequencies are determined observationally, we can calculate $\langle\Delta\nu\rangle$ as the mean difference between the frequencies of successive modes. The $\langle\Delta\nu\rangle$ value we calculated for this star is 53.4 $\mu$Hz. Another important aspect regarding sensitive models under asteroseismic constraints is selecting the value of $\langle\Delta\nu\rangle$. Since we calculated the $\langle\Delta\nu\rangle$ of the model from the differences between frequencies, it {is} more appropriate to use the $\langle\Delta\nu\rangle$ calculated {similarly} (53.4 $\mu$Hz). In the following section, we use both values of $\langle\Delta\nu\rangle$ as constraints for the interior models. 

\subsection{Effect of metallicity and $\chi^2$-method for modelling}
{Literature reports large differences} in the {parameter} values, especially, $T_{\rm eff}$, {determined from the spectral data}.
In the above models, while the $T_{\rm eff}$ values of the models are compatible with the observation, there is a significant difference between $Z_{\rm s}$. According to the observational [Fe/H] values given in Table \ref{tab:obs_nonasteroseismic}, $Z_{\rm s}$ ranges from 0.0074 to 0.0104. These values are significantly lower than {those} of the models in Table \ref{tab:ABCD_modeldata} ($Z_{\rm s} = 0.013 - 0.025$). {Hence, we aim to construct} a more suitable model.

Recently, many studies have explored the rate of the $^{14}$N(p,$\gamma$)$^{15}$O nuclear reaction. For the rate of this reaction, {\small MESA} has NACRE ({\tt set\_rate\_n14pg = 'NACRE'}) and JINA options. Previous models are constructed using JINA. Models constructed with the NACRE option are listed in Table \ref{tab:NACRE_JINA}. In NACRE \citep{Angulo1999} and JINA \citep{Cyburt2010}, the S-factor for this nuclear reaction was taken as $1.77 \pm 0.20$ (\cite{2001NuPhA.690..755A}) and $1.61 \pm 0.08$ (Imbriani et al. 2005), respectively.

Table \ref{tab:NACRE_JINA} {also lists} the surface chemical compositions and $\chi^2_{\rm seis}$ values of the models obtained with JINA and NACRE reaction rates for $^{14}$N(p,$\gamma$)$^{15}$O. The $\chi^2_{\rm seis}$ of the models are computed from
\begin{equation}
{{\chi^2_{\rm seis}}}={\frac{1}{3}}{\sum\limits_{i=1}^{3}{\left({\frac{{f_{i\rm obs}}-{f_{i\rm mod}}}{e_{i \rm obs}}}\right)^2}}
\label{eq:chisquare.seis}
\end{equation}
where ${f_{i\rm obs}}$ are the observed values of $\langle\Delta\nu\rangle$, 
$\nu_{{\rm min}0}$ and $\nu_{{\rm min}1}$. Their uncertainties and model values are represented by $e_{i\rm obs}$ and $f_{i\rm mod}$, respectively. 

{{When} constructing the interior models {listed} in Table \ref{tab:NACRE_JINA}, we {do not} calibrate the models over $M$, $R$ or $T_{\rm eff}$. We only consider the asteroseismic parameters in the computation of $\chi^2_{\rm seis}$ and check if the models with sufficiently small $\chi^2$ value have $Z_{\rm s}$ in good agreement with the observations. As a result of the analysis, $M$, $R$ and $T_{\rm eff}$ are given as output. In conclusion, three asteroseismic and one spectroscopic ($Z_{\rm s}$) constraints can be used to build a unique interior model for KIC 7747078.} %

These models are those with the smallest $\chi^2_{\rm seis}$ values among many constructed using $Z_{0} = 0.0115 - 0.0125$, $M = 1.15 - 1.20$, $\alpha = 1.8111 - 1.8511$, $Y_{0} = 0.2687 - 0.2723$ and $t_{9} = 4.6 - 6.0$. {The first four rows of Table \ref{tab:NACRE_JINA} show the} $\chi^2_{\rm seis}$ of the models computed by taking $\langle\Delta\nu\rangle = 53.40 \mu$Hz. The smallest $\chi^2_{\rm seis}$ value is for Model N1 obtained using $M = 1.17$ $\rm M_{\sun}$, $R = 1.975$ $\rm R_{\sun}$, $Z_{0} = 0.0121$, $Y_{0} = 0.2689$, and $\alpha = 1.8311$. The age of this model is 5.17 Gyr. {For this model,} $Z_{\rm s} = 0.0103$, {which} is in good agreement with the observational values. This is our favourite model with the NACRE rate for $^{14}$N(p,$\gamma$)$^{15}$O. For Model N2, where only $Y_{0}$ is slightly higher and the remaining parameters are the same, $\chi^2_{\rm seis}$ = 0.18. 
{The value of $\chi^2_{\rm seis}$ for the next model (Model N3) is 0.40.}
{This value is obtained} with a slightly greater value of $\alpha$ ($\alpha$ = 1.8511) compared to that of Model N2. 

The $Z_{\rm s}$ of models {listed} in Table \ref{tab:NACRE_JINA} is {approximately} 0.01. {This value} implies that $[{\rm Fe}/{\rm H}] = -0.127$, {which agrees} better with the observed values given in Table \ref{tab:obs_nonasteroseismic} than {the value} of Model A in Table \ref{tab:ABCD_modeldata}. 
The $Z_{\rm s}$ of Model N4 corresponds to $[{\rm Fe}/{\rm H}] = -0.140$, {which is} in perfect agreement with the {observed} value of [Fe/H] {reported} by \cite{Brewer2016}. The difference between Model N4 and Models N1 - N3 is the mixture of metallicity for the low-opacity temperature. For Model N4, the mixture is as {reported} by \cite{Asplund2009} ({\tt kap\_lowT\_prefix = 'lowT\_fa05\_a09p'}). The remaining models {contain} the mixture {reported by} \cite{GS98} (GS98, {\tt kap\_lowT\_prefix = 'lowT\_fa05\_gs98'}). 
Models N2 and N4 have {identical} $M$, $Z_{0}$, $Y_{0}$ and $\alpha$ values. The $Z_{\rm s}$ of Model N4 is slightly less than that of Model N2. {The $\chi^2_{\rm seis}$ value increases to 0.34 for Model N4.}

Models N5 $-$ N7 are constructed using $\langle\Delta\nu\rangle = 53.22$ $\mu$Hz. 
Among {them}, Model N6 shows the {lowest} $\chi^2_{\rm seis}$. {This model} is very similar to Model N1, {and} the mass difference between {them} is 0.01 $\rm M_{\sun}$. Models 5 and 7 have {identical} input parameters ($M$, $Y_{0}$ and $Z_{0}$) except $\alpha$, and {they provide similar} model results, including {the} ages. 
%
%
%
%

Model J, detailed in Table \ref{tab:NACRE_JINA}, is {constructed} with the {\tt set\_rate\_n14pg = 'jina reaclib'} option for $^{14}$N(p,$\gamma$)$^{15}$O, {for which} $M=1.160$ $\rm M_{\sun}$ gives the smallest value of $\chi^2_{\rm seis}$ as $0.10$. 

{The effects of different input parameters, such as mixtures of metallicity and nuclear reaction rate options, on frequencies have been tested. The models analysed are listed in Table \ref{tab:NACRE_JINA}.
Although the models are similar, the values of $\nu_{\rm min0}$ and $\nu_{\rm min1}$ vary between $1030 - 1050$ $\mu {\rm Hz}$ and $788 - 800$ $\mu {\rm Hz}$, respectively, causing differences in $\chi^2_{\rm seis}$. 
These changes in minima indicate how precise the method is.}

\subsection{Evolution grids for KIC 774708 and three-dimensional analysis}
\label{sec:grids}
{To verify the uniqueness of a model for KIC 7747078, we construct an evolution grid with varying $M$, $Z_0$ and age parameters. The grid ranges are 1.120-1.220 $\rm M_{\sun}$ for $M$, 0.0111-0.0131 for $Z_0$ and 5.12-5.22 Gyr for $t_9$. The grid steps are $\Delta{M} = 0.01 \rm M_{\sun}$, $\Delta{Z} = 0.0002$ and $\Delta{t_9} = 0.01 {\rm Gyr}$.}

\begin{figure}
    \centering   \includegraphics[width=\columnwidth]{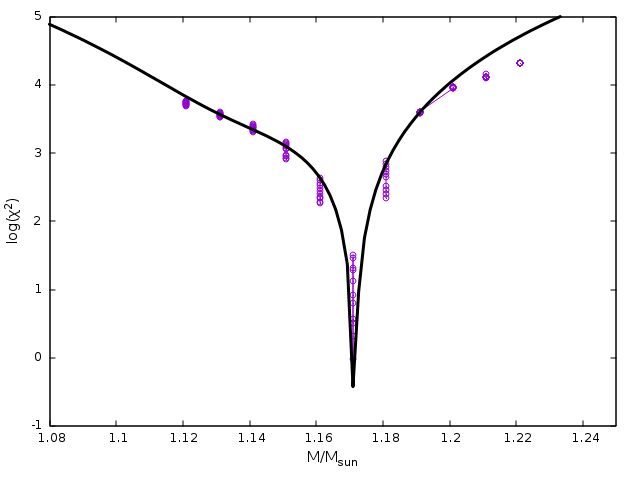} \includegraphics[width=\columnwidth]{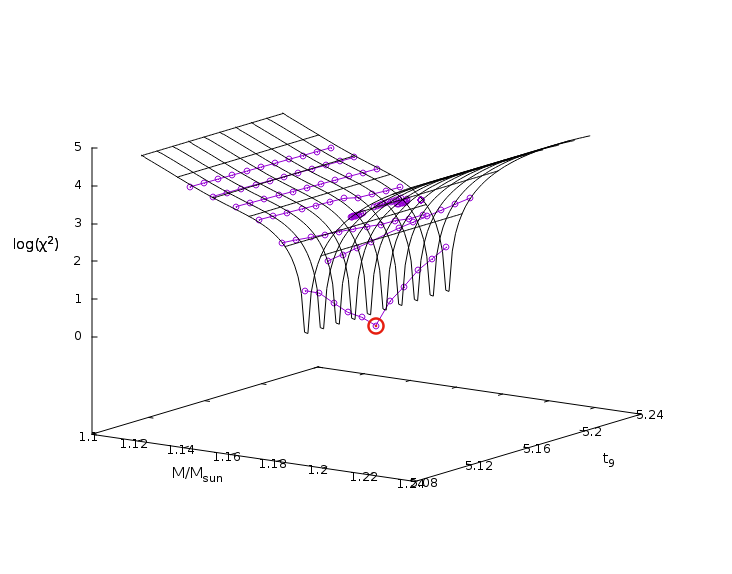} 
    \caption{{In the upper panel, logarithmic $\chi2$ is plotted against mass using the grid for KIC 7747078. $Z_0$ of the models is 0.0121. The lowest $\chi^2$ value is achieved at a mass of 1.171 $\rm M_{\sun}$.
    In the lower panel, logarithmic $\chi2$ is plotted against $M$ and $t_9$. $M$ and $t_9$ of the best model are 1.171 $\rm M_{\sun}$ and 5.165 Gyr, respectively. $t_9$ does not distinguish models at other masses. The best model is highlighted using a red circle.}}
    \label{fig:chi2Mt9}
\end{figure}
{Accurate mass determination is crucial for identifying the best model, as indicated by a very low $\chi2$ for a single value of $M$. The values $\nu_{\rm min1}$, $\nu_{\rm min0}$, $\Delta\nu$ and $Z_{\rm s}$ values are used to calculate $\chi2$ for the grids.}
{In the upper panel of Fig. \ref{fig:chi2Mt9}, $\log\chi2$ is plotted versus mass. The Z value of the models in the upper panel is 0.0121. The lowest $\chi2$ occurs at 1.171 $\rm M_{\sun}$, while masses other than this result in $\chi2$ values exceeding $100$. This result underscores the significance of accurate mass determination is reaching a unique model. In the lower panel of Fig. \ref{fig:chi2Mt9}, $\log\chi2$ is plotted against $M$ and $t_9$. Models with the lowest $\chi2$ values, particularly at $M = 1.171 M_{\sun}$, are differentiated by $t_9$, though no $t_9$ distinction is observed in other masses. At 5.165 Gyr, $\chi2$ is approximately 0.3. The best model we obtain with $M$ and $t_9$ is marked with a red circle in the graph.}

\begin{figure}
    \centering
    \includegraphics[width=\columnwidth]{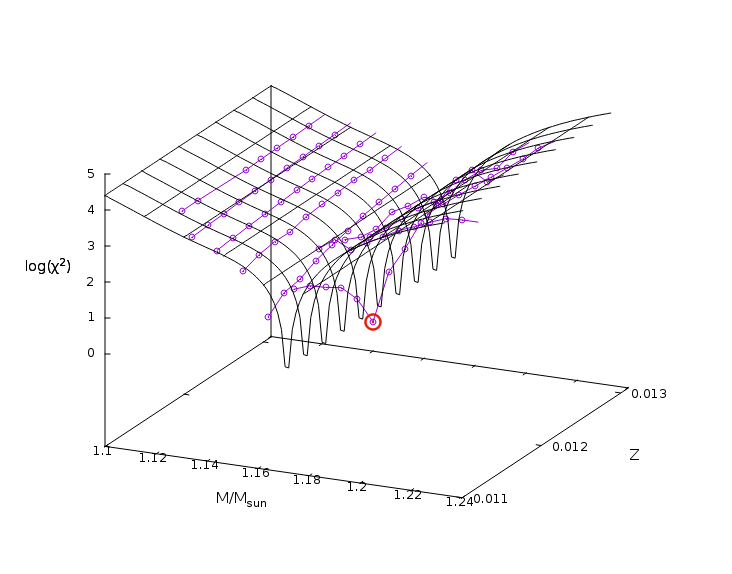}
    \caption{{Logarithmic $\chi2$ is plotted against $M$ and $Z_0$, showing the smallest $\chi2$ value at a certain $M$ and $Z_0$, indicating the unique model at that $M$. $M$ and $Z_0$ of the model are 1.171 $\rm M_{\sun}$ and 0.0121, respectively. $Z_0$ does not indicate another unique model in models with masses greater and less than 1.171 $\rm M_{\sun}$.
    The best model, as indicated by $Z_0$, is marked with a red circle.}}
    \label{fig:chi2MZ}
\end{figure}
{The second key parameter in stellar structure and evolution is $Z$. 
However, it is difficult to determine $Z$ observationally.
Asteroseismic modelling method can be used to estimate $Z$ values. This study examines how $Z_0$ affects frequencies using the grid. In Fig. \ref{fig:chi2MZ}, $\log\chi2$ is plotted against $M$ and $Z_0$, revealing a single solution where $\chi2$ is minimised using asteroseismic parameters. The lowest $\chi2$ value is achieved at $Z_0 = 0.0121$ for models with masses of 1.171 $\rm M_{\sun}$. Other $Z_0$ values do not yield a specific model for above or below 1.171 $\rm M_{\sun}$. The best model is drawn with a red circle in Fig. \ref{fig:chi2MZ}.} 

{Using the grids, we derive several models with different fundamental parameters for KIC7747078. Upon examining the asteroseismic data, a unique model, referred to as Model N1, is identified. In this case, the $M$, $Z_0$ and $t_9$ values of the unique model with the lowest $\chi2$ are 1.171 $\rm M_{\sun}$, 0.0121 and 5.165 Gyr, respectively.}

\subsection{Analysing mixed and radial modes of KIC 7747078 using the astero module of MESA}
\label{sec:mixmode_astero}
{As stars evolve, some oscillation modes are mixed, with the mixed mode frequencies providing insights into the core structure of stars \citep{Deheuv2012, Beck2012}. These frequencies exhibit p-mode features in the envelope and g-mode features in the core. Since there is no $l = 0$ mode in the g-mode, radial mode frequencies remain unaffected.} {Analysis of the observational $l = 1$ and $l = 2$ mode frequencies of the subgiant KIC 7747078 that $l = 1$ modes are strongly affected, while $l = 2$ modes are either unaffected or minimally affected.}

\begin{figure}
    \centering
    \includegraphics[width=\columnwidth]{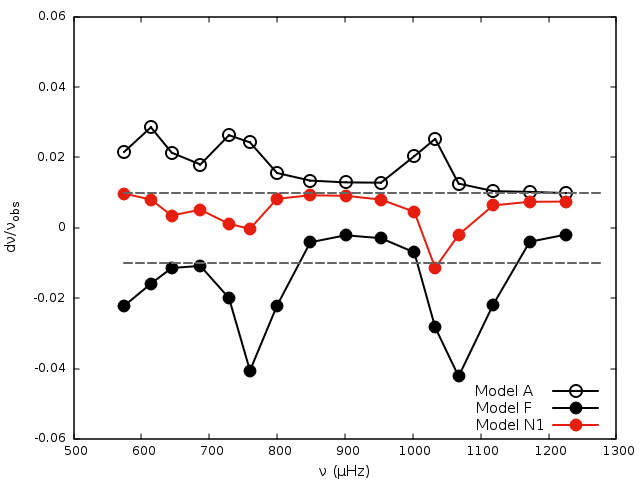}    \includegraphics[width=\columnwidth]{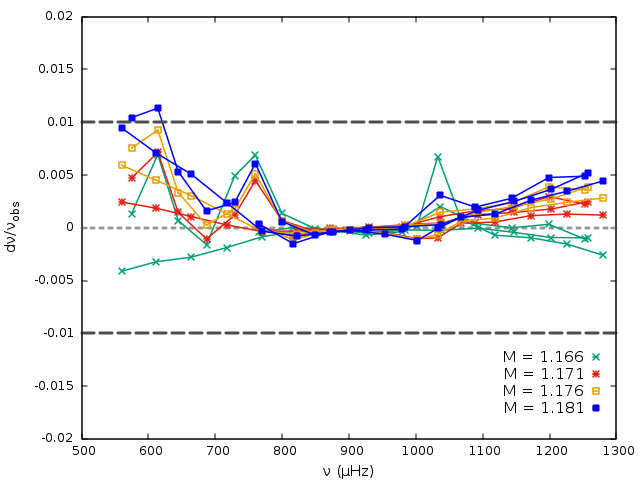}
    \caption{The fractional difference ($d\nu/\nu_{\rm obs}$) of the frequencies of the three models for the mode $l = 1$ is plotted against $\nu$ in the top panel. The dashed lines show the fractional difference of 0.01. Among the three models, N1 is the model that best matches the observation. In the bottom panel, fractional differences of the models constructed with the astero module for the modes $l =$ 0, 1 and 2 are given. The most compatible models have a mass of 1.166 and 1.171 $\rm M_{\sun}$.}
    \label{fig:AEN1_obs-frek}
\end{figure}
{The fractional difference of the $l = 1$ mode frequencies of model A and N1, which are good in terms of minima, and model E, which is good in terms of frequencies, with the observation is given in Fig. \ref{fig:AEN1_obs-frek}. 
Model N1 aligns with observational frequencies within a 0.01 fractional difference, whereas Models A and E exhibit differences up to 0.04. Among these three models, Model N1 best matches the mixed modes.} 

\begin{figure}
    \centering
    \includegraphics[width=\columnwidth]{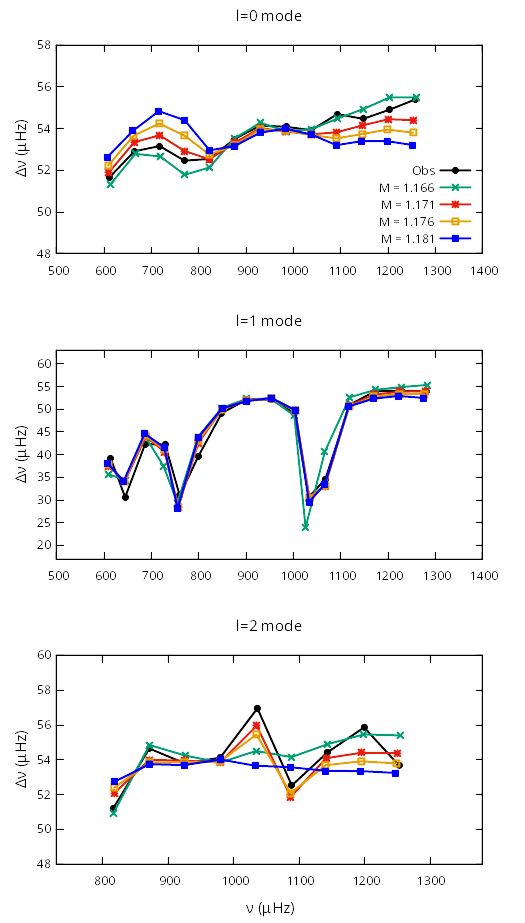}
    \caption{{$\Delta\nu$ against $\nu$ plots for modes $l =$ 0, 1 and 2 of models constructed using the astero module. The top panel shows the $l = 0$ mode, the middle panel the $l = 1$ mode and the bottom panel the $l = 2$ mode. While the model frequencies can be distinguished in modes $l = 0$ and $l = 2$, the models cannot be distinguished from each other in the $l = 1$ mode. The 1.166 and 1.171 $\rm M_{\sun}$ models in the $l = 0$ mode and the 1.171 and 1.176 $\rm M_{\sun}$ models in the $l = 2$ mode align well with the observational frequencies. The $\chi2_{\rm seis}$ value calculated from $\nu_{\rm min1}$, $\nu_{\rm min0}$ and $\Delta\nu$ determined from the $l = 0$ mode indicates the model with a mass of 1.171 $\rm M_{\sun}$ as the unique model.}}
    \label{fig:astero_Dnu-nu}
\end{figure}
{
Using the parameters of the unique model N1, a new model is developed using the astero module of {\small MESA}. The results of the best model obtained from this module are presented as Model AN in the last row of Table \ref{tab:NACRE_JINA}.} 
For this purpose, four models with $Z_s=$0.0100 were constructed in the range of 1.166 – 1.181 $\rm M_{\sun}$, with increments of 0.005 $\rm M_{\sun}$. These models are run until the radius provides the best match for the mode frequencies with $l =$ 0, 1 and 2.
The frequency values from all four models are aligned well with observations.
The fractional difference between the model and the observed individual frequencies varies between -0.4 to 1.1 percent. The fractional differences for these astero models are shown in the bottom panel of Fig. \ref{fig:AEN1_obs-frek}. All models agree well with the observations at frequencies between 800 and 1000 $\mu {\rm Hz}$ for modes $l=$ 0, 1 and 2. However, the differences increase at frequencies above 1000 $\mu {\rm Hz}$ and below 800 $\mu {\rm Hz}$. When the individual frequencies of the modes $l=$ 0, 1 and 2 are examined together, the models with a mass of $M=1.166$ $\rm M_{\sun}$ and $M = 1.171$ $\rm M_{\sun}$ give the lowest mean fractional difference.

{The mixed mode frequencies with $l = 1$ align well with observational frequencies in the $\Delta\nu - \nu$ plot (middle panel in Fig. \ref{fig:astero_Dnu-nu}). 
For the $l=2$ mode, two models, namely those with masses 1.171 $\rm M_{\sun}$ and 1.176 $\rm M_{\sun}$, fit zigzag structure seen in the $\Delta\nu - \nu$ plot (bottom panel in Fig. \ref{fig:astero_Dnu-nu}). This zigzag pattern at high frequencies becomes more linear at both lower and higher masses.
Distinct patterns are observed for the $l = 0$ modes (top panel in Fig. \ref{fig:astero_Dnu-nu}), where the successive difference between frequencies increases at frequencies below approximately 900 $\mu {\rm Hz}$ and decreases above it as mass increases.  
The $\chi2_{\rm seis}$ values, calculated using $\Delta\nu$, $\nu_{\rm min1}$ and $\nu_{\rm min0}$, are 0.83 and 0.34 for the 1.166 $\rm M_{\sun}$ and 1.171 $\rm M_{\sun}$ models, respectively. The other two models have $\chi2_{\rm seis}$ values greater than 2.
The unique model is Model AN (last row of Table \ref{tab:NACRE_JINA}) with a mass of 1.171 $\rm M_{\sun}$, which has the lowest $\chi2_{\rm seis}$ value of the four models.} 

The uncertainties in $M$, $R$ and age are computed using the Monte Carlo simulation method and are found as 0.019 $\rm M_{\sun}$, 0.011 $\rm R_{\sun}$ and 0.28 Gyr, respectively.

{Models N1 and AN exhibit very similar parameters, with the same $M$ and $Z_0$. The differences in $T_{\rm eff}$, $Y_0$ and age are minimal. However, Model AN is identified as the best and unique model when individual frequencies are considered.}

\section{Conclusions}
In this study, we analyze the {SG star} KIC 7747078, {which was} observed by the Kepler space telescope. The {aim is} to obtain a unique interior model representing the star using ${\nu_{\rm min}}$s and individual frequencies. 
We use ${\nu_{\rm min}}$ because the asteroseismic parameters such as ${\Delta\nu}$ and ${\nu_{\rm max}}$ are related to the general properties of the star (e.g. $\rho$ {and} $\log{g}$). {Therefore, stars with different masses and radii can have similar ${\Delta\nu}$ and ${\nu_{\rm max}}$. ${\delta\nu_{\rm 02}}$ is not a good indicator {of the age of post-MS} stars {because a} change in the internal structure of an evolved star reduces the sensitivity of this parameter to the core region. {Hence}, asteroseismic parameters such as ${\Delta\nu}$, ${\nu_{\rm max}}$ and ${\delta\nu_{\rm 02}}$ cannot be used to distinguish models for evolved stars}.

The first step of the study is to estimate the mass and radius of the star using different methods (Section $\ref{sec:MRcomput}$). The discrepancy {among} the {previously reported} data {of the star} makes it difficult to determine its fundamental parameters. This discrepancy {also} makes it difficult to {construct a} unique model of the star. 

{
The interior models are constructed with the star and astero modules in the MESA evolution code. In our search for a unique model with a vast parameter pool, we mainly use the star module, which applies the standard \cite{Kjeldsen2008} correction for near-surface effects on oscillation frequencies. Once we obtain a unique model, we construct new models in the astero package by taking the parameters of this model as initial values and using the individual oscillation frequencies, and again obtain a unique model. In this module, we use the \cite{2014A&A...568A.123B} method to account for near-surface effects. The differences between the unique models obtained with the two modules are minimal.
}

{For the interior models constructed using the star module, the initial model values are first calculated based on scaling relations.} 
Each model shows a unique evolutionary track and age (see Fig. \ref{fig:evrim_modelC-D}). To distinguish them, the frequency pattern is first examined. Since the frequencies of the models are intertwined in the ${\Delta\nu} - {\nu}$ graph, {the models cannot be} distinguished in this way. Therefore, the minimum and individual frequencies of the models are compared with observational values. {Model A is obtained as} the common solution as given {per the} $\nu_{\rm min1}$ and $\nu_{\rm min0}$ {values of} the models we constructed with $\alpha_{\sun}$. 
{The difference between} the frequencies of {Model A and} the observational frequency {is} ~11 $\mu$Hz, {and the} difference between the individual frequencies decreases with increasing $\alpha$ (Model D).

Two solutions are obtained using{($B-V$) and ($V-K$) colours and by comparing $M_{\pi}$ and $ M_{\rm sca}$}. Models E and F show the results obtained using these solutions. Model E {provides} the best agreement with observational frequencies. However, the minima shift to higher frequencies.

{Among the six interior models of KIC 7747078 given in Table \ref{tab:ABCD_modeldata},  the reference frequencies of Model A and the individual frequencies of Model E are in very good agreement with the observations.} The basic properties of Model A are $M = 1.208$ $\rm M_{\sun}$, $R = 2.00$ $\rm R_{\sun}$ and age $t = 5.24$ Gyr. 
For Model E, $M = 1.279$ $\rm M_{\sun}$, $R = 2.02$ $\rm R_{\sun}$ and $t = 4.04$ Gyr. These ages are significantly less than {those reported} by \citet{Chaplin2014} and in good agreement with the ages found by \citet{Yildiz2019}. 

While the $\nu_{\rm min0}$ and $\nu_{\rm min1}$ of Model A are in very agreement with the observations, the observational and model oscillation frequencies {differ by 11 $\mu$Hz}. For Model D, {which has} a greater value of $\alpha$ ($\alpha=2.454$) {than Model A,} the difference {in frequencies} reduces to about 4 $\mu$Hz. However, the surface metallicity of these models {does} not agree with {the observed spectral data}.

We apply the $\chi^2$-method to {adopt} the asteroseismic constraints. The three constraints of the models are the observed values of $\langle\Delta\nu\rangle$, 
$\nu_{{\rm min}0}$ and $\nu_{{\rm min}1}$. 
According to the $\chi^2_{\rm seis}$ value, for $\langle\Delta\nu\rangle$ = 53.4 $\mu$Hz, the best fitting model is Model N1 with $\chi^2_{\rm seis}=0.02$. The basic parameters of {Model N1 are as follows}: $M=$1.17 $\rm M_{\sun}$, $R=$1.975 $\rm R_{\sun}$, $T_{\rm eff}=$5977 K and $t_9=5.17$ Gyr. All the models {listed} in Table 3 with slightly different input parameters {($M$, $R$, $Y_0$ and $\alpha$)} are {somewhat} similar to Model N1. {That is}, the use of the three asteroseismic parameters and surface metallicity as constraints yields a unique interior model for KIC 7747078.

{
We demonstrate that it is feasible to identify a unique model of an evolved star using $\nu_{{\rm min}1}$, $\nu_{{\rm min}0}$, $\Delta\nu$ and $Z_s$. However, the mass needs to be obtained precisely. This is evident in the grids for KIC 7747078, where the minimum $\chi^2$ value is observed at $M =$ 1.17 $\rm M_{\sun}$ for all combinations of $M$, $Z_0$ and $t_9$. The unique model with mass $1.171 \rm M_{\sun}$ in the grid is strongly pointed with $Z_0 = 0.0121$ and $t_9 = 5.165$ Gyr.
}

{In discussing a subgiant star’s unique model, we examined the mode frequencies for $l =$ 0, 1 and 2 (Fig. \ref{fig:astero_Dnu-nu}). For this, we construct interior models using astero module of MESA with initial parameters the same as that of Model N1. This module attempts to fit the model frequencies to the observed frequencies. 
In models with different masses, changes in $\nu_{\rm min1}$ and $\nu_{\rm min0}$ determined from $l =$ 0 frequencies and different patterns in the $\Delta\nu - \nu$ graph led us to a single solution. However, the mixed modes with $l =$ 1 appear unaffected by mass changes, while $l =$ 2 frequencies show some mass-dependent variation, albeit less pronounced than $l =$ 0. Therefore, $l =$ 0 mode frequencies are found to be more effective in identifying a unique model. Among the four models, the model with $M = 1.171 \pm 0.019$ $\rm M_{\sun}$ (Model AN) shows the lowest $\chi^2$. Its age and initial helium abundance are $5.15 \pm 0.29$ Gyr and 0.2689, respectively. 
This is the unique model of KIC 7747078.} 

{Despite the shallowness of $\nu_{\rm min0}$ for KIC 7747078, a unique model is derived using the $\chi^2$ method, incorporating asteroseismic constraints and reference frequencies. To achieve a unique solution using both $\nu_{\rm min1}$ and $\nu_{\rm min0}$, it would be useful to analyse other Kepler Legacy stars that have at least two reference frequencies.}

\section*{Acknowledgements}
This work is supported by the Scientific and Technological Research Council of Turkey (TÜBİTAK: 123F019). We are grateful to Ege University Planning and Monitoring Coordination of Organizational Development and Directorate of Library and Documentation for their support in editing and proofreading service of this study.


\section*{Data Availability}

 The data underlying this article will be shared on reasonable request to the corresponding author.
 










\bsp	
\label{lastpage}
\end{document}